\documentclass[twoside]{ilcws08}
\usepackage[latin1]{inputenc}
\usepackage[dvips]{graphicx,epsfig,color}
\usepackage{wrapfig,rotating}
\usepackage{amssymb,amsmath,array}

\begin{document}
\title{
Analysis of ZH recoil mass} 
\author{Kazutoshi Ito$^1$
\vspace{.3cm}\\
1- Department of Physics, Tohoku University \\
Research Center for Neutrino Science, Tohoku University, Sendai , Japan
}

\maketitle

\begin{abstract}
 The precise measurement of the Higgs mass is the most important program
 at International Linear Collider (ILC).
 Using
 $e^+e^-\rightarrow ZH$ process, the
 mass of Higgs boson can be
 measured by two lepton tracks from decay of Z boson, even if the
 Higgs boson decays into invisible particles.
 We report the measurement accuracy on the Higgs recoil mass
 and the cross sections for $ZH\rightarrow \ell^+\ell^-X$ at
 $\sqrt{s}$=250 GeV with the integrated luminosity of 250 fb$^{-1}$.
\end{abstract}

\section{Introduction}
The Higgs-strahlung $e^+e^- \rightarrow ZH$, is the most important mode
to study the Higgs mass, branching ratio, etc. at the ILC.
Especially, the leptonic decay mode of Z boson realizes the precise
measurement of the Higgs mass due to the finite center of mass energy at
ILC, for example, 
the Higgs mass can be measured as the recoil mass against the Z
boson ($m_{recoil}$) by using reconstructed mass and energy of a Z
 boson as followings;
\begin{eqnarray}
 m_{recoil}^2 = s + m^2_{Z} - 2\cdot E_{Z}\cdot \sqrt{s},
\end{eqnarray}
where $\sqrt{s}$ is the center of mass energy, and $m_{Z}$ and
$E_{Z}$ are the mass and energy of a Z boson
reconstructed by the two lepton
tracks. Since the decay products from Higgs bosons are not used in this
study, the Higgs mass can be
measured even if the Higgs boson decays into invisible particles.
In this paper, we
report our analysis status on measurement accuracy of the Higgs recoil
mass and the cross 
section of $e^+e^-\rightarrow ZH$ by using 
$ZH\rightarrow e^+e^-/\mu^+\mu^-X$.
 
\section{Simulation}
In this study, we used the geometry of
 LDC$^\prime$ detector model (\verb$LDCPrime_02Sc$), which is prepared
 for the detector optimization study for ILD. We assumed the mass of Higgs boson
 as 120 GeV.
The center of mass energy was
set to 250 GeV, where
the initial beam spread was considered to be 0.28\% for the electron
 beam and 0.18\% for the positron beam.
The beam simulation was done by CAIN \cite{CAIN} with
the initial and final state radiation (ISR and FSR) and beamstrahlung.
WHIZARD \cite{whizard} was used as the event generator, and hadronization was done by
Pythia 6.409 \cite{pythia}.
The generated events were simulated by Mokka and reconstruction was done
by Marlin \cite{ilcsoft}.

We considered only 
$ZZ \rightarrow e^+e^-X/\mu^+\mu^-X$ events as
background. The cross section of the
signal events were 7.5 fb and these of ZZ events were 78.7 fb for
$ZZ\rightarrow e^+e^-X$ and 79.0 fb for $ZZ\rightarrow \mu^+\mu^-X$.
In the analysis, the number of events
was scaled to 250 fb$^{-1}$.

\section{Analysis}
\begin{table}[htbp]
\begin{center}
 \begin{tabular}{|l|r|r|}
  \hline
  \multicolumn{3}{|c|}{$ZH \rightarrow eeX$} \\\hline
  & signal& $ZZ\rightarrow eeX$\\\hline
  No cut& 1923(1.00)& 19685(1.00)\\ \hline
  2-track selection& 1482(0.771)& 11931(0.606)\\ \hline
  85 $<M_Z<$ 97 GeV& 1065(0.554)& 7844(0.399)\\ \hline
  $|\cos\theta_{lepton}|<$ 0.95& 967(0.503)& 7140(0.363)\\ \hline
  $|\cos\theta_{Z}<$ 0.9& 891(0.463)& 5657(0.287)\\ \hline
  \hline
  \multicolumn{3}{|c|}{$ZH \rightarrow \mu\mu X$} \\\hline
  & signal& $ZZ\rightarrow \mu\mu X$\\\hline
  No cut& 1923(1.00)& 19685(1.00)\\ \hline
  2-track selection& 1766(0.918)& 13799(0.701)\\ \hline
  85 $<M_Z<$ 97 GeV& 1530(0.796)& 10856(0.552)\\ \hline
  $|\cos\theta_{lepton}|<$ 0.95& 1382(0.718)& 9831(0.499)\\ \hline
  $|\cos\theta_{Z}|<$ 0.9& 1266(0.658)& 7659(0.389)\\ \hline
 \end{tabular}
\end{center}
 \caption{Reduction summary. The number of events was scaled to 250
 fb$^{-1}$.}
 \label{tab:reduction}
\end{table}

\begin{wrapfigure}{r}{0.5\columnwidth}
\centerline{\includegraphics[width=0.45\columnwidth]{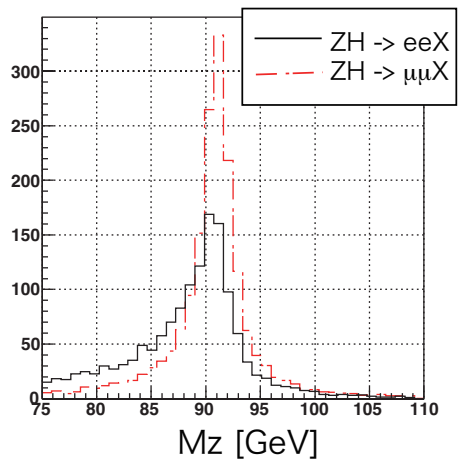}}
\caption{The distribution of the reconstructed Z mass for
 $ZH\rightarrow ee/\mu\mu X$.}\label{Fig:zmass}
\end{wrapfigure}

To identify electrons or muons coming from decay of Z bosons,
we reconstructed the invariant mass by using two charged
tracks with track energy above 10 GeV.
Then, a pair of the lepton tracks which had the nearest mass to the Z boson was
selected.
Figure \ref{Fig:zmass} shows the distribution of the reconstructed Z
mass. $ZH\rightarrow eeX$ events have a wider distribution than the
$ZH\rightarrow\mu\mu X$ events because electrons emit the bremsstrahlung
photons in the tracker.
To use well
reconstructed events, we applied the selection cut of
$85 < M_Z < 97$ GeV for the signal and background events.

We applied the angular cut for the lepton tracks and reconstructed Z bosons.
At first,
tracks with $|\cos\theta_{lepton}|<$ 0.95 was selected
since the coverage of TPC is $|\cos\theta| <$ 0.98.
$e^+e^- \rightarrow ZZ$ events are the t-channel process,
therefore, the Z boson tends to be generated in forward-backward region.
On the other hand, 
the Z boson is generated uniformly for the signal events
because $ee \rightarrow ZH$ events come from the s-channel process and
the Higgs boson is a scalar particle.
We required the angle of the reconstructed Z boson to be
$|\cos\theta_{Z}|<$ 0.9.

Table \ref{tab:reduction} shows the reduction summary at each selection cut.
After all the cuts, the selection efficiency was 46.3\% for
$ZH \rightarrow e^+e^-X$ and 65.8\% for
$ZH \rightarrow \mu\mu X$.
On the other hand, the
efficiency for the background events was 28.7\% and 38.9\% for
$ZZ\rightarrow e^+e^-X$ and $ZZ\rightarrow \mu^+\mu^-X$, 
respectively. 

\section{Results}
\begin{figure}[htbp]
\begin{center}
 \includegraphics[width=\hsize]{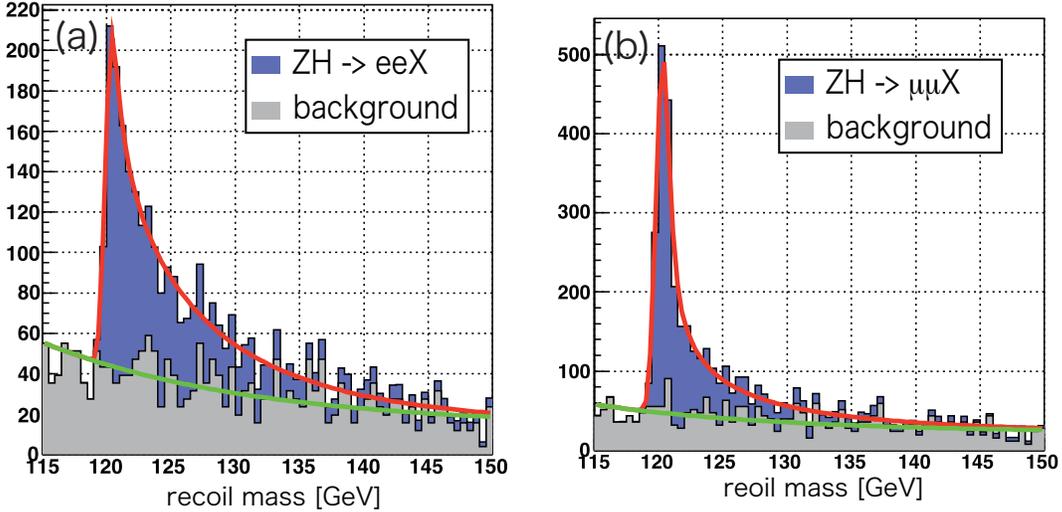}
 \caption{Distribution of the reconstructed Higgs recoil mass for
 $ZH\rightarrow eeX$ (a) and $ZH\rightarrow \mu\mu X$ (b).}
 \label{Fig:fit}
\end{center}
\end{figure}

Figure \ref{Fig:fit}
shows the distributions of the reconstructed Higgs recoil mass
for $ZH\rightarrow e^+e^-/\mu^+\mu^-X$.
A peak can be seen around 120
GeV in both distributions.
To obtain the Higgs recoil and cross section, the distributions were
fitted by the empirical function as follows \cite{DOD};
\begin{eqnarray}
 F(m) &=& N_{sig}e^{-Am}\int F_H(m+t)e^{-\frac{t^2}{2\sigma^2}}dt +
  F_Z(m), \nonumber \\
 F_H(m) &=& \left(\frac{m-M_H}{\sqrt{s}-M_H}\right)^{\beta -1},
\end{eqnarray}
where $M_H$ is the mass of Higgs boson, $\sqrt{s}$ is the center of mass
energy, and $N_{sig}$ is a normalization factor for the signal events.
$F_H(m)$ is a function to take into account of the effect of the
bremsstrahlung, which is convoluted with the Gaussian function,
and $e^{-Am}$ is a correction term. $F_Z(m)$ is a exponential
function to fit the background distribution.
The parameters of
$F_Z(m)$ except for the normalization factor were determined by using the
the independent background samples.

From the fitting results, we obtained the mass of Higgs boson
as $120.0 \pm 0.10$ GeV and $120.1 \pm 0.041$ GeV for
$ZH\rightarrow e^+e^-X$ and $ZH\rightarrow \mu^+\mu^-X$, respectively.
The cross section can be calculated from the integral of the fitted
function.
 The measured cross section of $ZH \rightarrow
e^+e^-X$ was $7.5\pm 0.35$ fb and that of $ZH
\rightarrow \mu^+\mu^- X$ was $7.7 \pm 0.29$ fb,
 which correspond to measurement accuracy of 
4.7\% and 3.8\%, respectively.

\section{Conclusions}
$e^+e^-\rightarrow ZH$ process is the golden mode to determine the Higgs
mass in the ILC, because the Higgs mass can be measured by the recoil
mass against the Z boson without any assumptions of the Higgs decay
modes.
We analyzed the recoil mass at $\sqrt{s}=$250 GeV with 250 fb$^{-1}$ for
the Higgs mass of 120 GeV, where $ee\rightarrow ZZ$ is considered as a
background. 
The Higgs boson mass was measured as $120.0 \pm 0.10$ GeV and $120.1
\pm 0.041$ GeV for $ZH\rightarrow e^+e^-X$ and $ZH\rightarrow
\mu^+\mu^-X$, respectively.
The cross section of $e^+e^- \rightarrow ZH \rightarrow
e^+e^-X$ was $7.5\pm 0.35$ fb (4.7\%) and that of $e^+e^- \rightarrow ZH
\rightarrow \mu^+\mu^- X$ was $7.7 \pm 0.29$ fb (3.8\%).

Although we included only $ZZ\rightarrow \ell^+\ell^-X$ events as
background events, other physics processes also become backgrounds.
For example, $ee\rightarrow e^+e^-/\mu^+\mu^-$ and $WW\rightarrow
e\nu_ee\nu_e$ contaminate in the signal region.
The next step, therefor, is to include these background events.

\section*{Acknowledgments}
I would like to thank Akiya Miyamoto for event generation, DESY
Production Group for detector simulation and reconstruction,
and all the member of the JLC-Software group
for useful discussions and helps. This work is supported in part by the
Creative Scientific Research Grant (No. 18GS0202) of Japan Society for
Promotion of Science and the JSPS Core University Program.


\begin{footnotesize}



%

\end{footnotesize}


\end{document}